\documentclass[12pt]{article}
\usepackage[symbol]{footmisc}
\def\correspondingauthor{\footnote{Corresponding author.  }}
\usepackage[left=2cm,top=2.5cm,right=2cm,bottom=2.5cm]{geometry}
\usepackage{amsmath, amssymb}

\usepackage{graphicx}
\usepackage{graphics}
\usepackage{epstopdf}
\usepackage{subfigure}
\usepackage{amsfonts}
\usepackage{sectsty}
\usepackage{sectsty}
\usepackage{hyperref}
\usepackage{cite}

\begin{document}
	
	\begin{center}
	\large{\bf{Wormhole Modeling Supported by Non-Exotic Matter}} \\
	\vspace{5mm}
	\normalsize{Gauranga C. Samanta$^1$ and Nisha Godani$^{2} {}$ \correspondingauthor{} }\\
	\normalsize{$^1$ Department of Mathematics, BITS Pilani K K Birla Goa Campus, Goa, India}\\
	$^2$ Department of Mathematics, Institute of Applied Sciences and Humanities\\ GLA University, Mathura, Uttar Pradesh, India\\
	\normalsize {gauranga81@gmail.com\\nishagodani.dei@gmail.com}
\end{center}
\begin{abstract}
	In the present paper, the modelling of traversale wormholes, proposed by Morris \& Thorne \cite{morris1}, is performed within the $f(R)$ gravity with particular viable case $f(R)=R-\mu R_c\Big(\frac{R}{R_c}\Big)^p$, where $\mu, R_c>0$  and $0<p<1$. The energy conditions are analyzed using the shape function $b(r)=\frac{r\log(r+1)}{\log(r_0+1)}$ defined by Godani and Samanta \cite{godani} and geometric nature of wormholes is analyzed. \end{abstract}

\section{Literature Survey}
The study of Morris-Thorne wormholes has become an important topic of study that first came into existence when Morris \& Thorne \cite{morris1} using general relativity put efforts to investigate the possibility of travel for  human being. Flamm \cite{flamm} first attempted to find out a solution of Einstein's field equations  that allows the travel in \& out of the throat without any restriction. But later, it was found to be unstable. Then Einstein \& Rosen \cite{eins-ros} performed an investigation of wormhole solutions in detail and obtained a solution which is known as Einstein Rosen bridge. Morris-Thorne wormholes or traversable wormholes are hypothetical tunnel like structures in space-time which are supported by an exotic matter that violates the null energy condition (NEC).
However, NEC is satisfied by classical forms of matter \cite{Hawking}.
In literature, wormhole geometries are  studied to find the configurations where the  weak energy condition is not violated.
Kar \cite{Kar} found the existence of static wormhole geometries such that the weak energy condition is not violated and discussed their properties.
Wang \& Letelier \cite{Wang} obtained non-static wormhole solutions  that satisfy weak and dominated energy conditions and consequently require less exotic matter in comparison of static wormhole solutions.
Kar \& Sahdev \cite{Sahdev} explored Lorentzian wormholes with a matter that satisfies energy conditions. They used several scale factors and found various results for Kaluza Klein and exponential inflation.
It is also observed that the energy conditions are violated in the presence of exotic matter that supports wormhole geometry \cite{Visser1}.
Roman \cite{Roman} investigated the possibility for the enlargement  of Lorentzian wormholes to macroscopic level through inflation. They defined a metric embedded in flat de-Sitter space and studied its properties.
Poisson and Visser \cite{Poisson} considered linear  thin-shell wormholes and studied linear perturbations around static wormhole solutions.
Barcel\'{o} and Visser \cite{Barcelo} considered massless conformally
coupled scalar fields and studied traversable wormholes.
Gonz\'{a}lez-D\'{i}az \cite{Gonz} studied the impact of accelerated expansion of the universe on shape and size of wormholes and ringholes.
Visser et al. \cite{Visser} obtained the existence of traversable wormholes in presence of very  small  amount  of  exotic  matter.
Gonz\'{a}lez-D\'{i}az \cite{Gonz1} studied the effect of accretion of phantom and dark energies on traversable wormholes.
Sushkov \cite{Sushkov} studied static and spherically symmetric wormholes and found their existence to be supported by phantom energy.
Lobo \cite{Lobo} examined the properties of traversable wormholes and investigated some particular wormhole structures.
Lobo \cite{Lobo2} studied traversable wormholes with phantom energy using the equation of state $p=\omega\rho$, where $\omega<-1$ and analyzed their stability.
B\"{o}hmer et al. \cite{Bohmer1} obtained exact wormhole solutions by using linear relationship between energy density and pressure and assuming different choices of form function.
Dotti et al. \cite{Dotti} explored static wormhole solution in dimensions $2n+1$, where $n$ is a natural number and $n\geq 2$.
For five dimension, they used Einstein-Gauss-Bonnet action and for higher dimensions, they used  Lovelock action.
Bronnikov and Krechet \cite{Bronnikov} obtained  wormhole solutions having asymptotically flat regions. They joined wormhole solutions to flat asymptotic regions and found a configuration consisting of  three regions (one internal and two externals).
Forghani et al. \cite{Forghani} developed asymmetric thin-shell wormholes using different redshift functions and found the weak energy condition to be satisfied.
Wormholes have also been explored in several other aspects \cite{Cataldo, Heydarzade, Moradpour,  El-Nabulsi}.

During last few years, the cosmologists have focused on the generalization of general relativity to modified theories of gravity to explain several phenomenons \cite{ Capozziello:2002rd, Capozziello:2003tk, Nojiri, Faraoni, Hu, Faul, Zhang, Cap, Sawicki, Amen, samanta1, samanta2, samanta3, moreas, moreas1, yousaf, Shabini, samanta4, samanta5, Nisha1, Nisha2}. The $f(R)$ theory of gravity is one among these generalized theories which has been used for the description of flat rotation curves of galaxies, cosmic  expansion of universe, formation of wormholes etc. \cite{Amendola, Nojiri1, Felice, Capozziello, Bamba, Martino}.
Cognola et al. \cite{Cognola} considered one and two steps models in $f(R)$ gravity in which both early inflation and late time acceleration arise in a natural way.
Linder \cite{Linder} used exponential $f(R)$ gravity model to describe the cosmic acceleration.
Amendoal and Tsujikawa \cite{Dark} discussed various significant $f(R)$ gravity models.
Nojiri et al. \cite{Nojiri4} reviewed modified gravity theories and discussed latest progress in various aspects such as  cosmology,  bouncing cosmology, late-time acceleration era etc.

The wormhole geometries are investigated in the background of $f(R)$ theory.
Nojiri et al \cite{Nojiri5} used effective action method and showed the existence of primordial wormholes at the early universe.
Further, Nojiri et al. \cite{Nojiri6} proved that for at least some initial conditions GUTs primordial wormholes are induced at the early universe.

Lobo and Oliveira \cite{Lobo1} studied traversable wormholes in $f(R)$ gravity  and found the higher order curvature terms in stress energy tensor to be responsible for the violation of null energy condition. They also found exact solutions for different choices of shape function and equation of state.
Saiedi and Esfahani \cite{Saiedi1} used power law expansion and a particular shape function and obtained exact wormhole solutions in $f(R)$ gravity. They considered traceless matter and obtained the satisfaction of both null and weak energy conditions.
Saiedi \cite{Saiedi} explored thermodynamic features of  evolving Lorentzian wormholes and developed wormholes using $f(R)$ model in radiation era.
Bahamonde et al. \cite{Bahamonde} studied wormhole solutions in $f(R)$ gravity and constructed dynamical wormholes approaching towards FLRW universe asymptotically.
Kuhfittig \cite{peter} derived some shape functions for some $f(R)$ functions and vice versa and obtained  wormhole solutions in $f(R)$ gravity.
Godani and Samanta \cite{godani} obtained wormhole solutions  with phantom energy by considering shape functions of two types. Recently,
Samanta et al. \cite{godani1} introduced a shape function and carried out a comparative study of traversable wormholes in $f(R)$ and $f(R,T)$ theories.

In the light of the above discussion, there are several choices for the selection of an $f(R)$ gravity model. We have chosen $f(R)=R-\mu R_c\Big(\frac{R}{R_c}\Big)^p$, where $\mu, R_c>0$  and $0<p<1$, which fulfills various conditions discussed later. Further, the  specific form of shape function $b(r)=\frac{r\log(r+1)}{\log(r_0+1)}$ defined by Godani and Samanta \cite{godani} is used. The aim of this work is to find a wormhole solution within the framework of modified $f(R)$ gravity and  explore the spherical regions where all energy conditions are satisfied.
The section-wise description of the work is as follows:  Field equations and
wormhole geometry is presented in section-2. In section-3, solutions of the wormhole geometry are presented.
In section-4, various energy conditions and results are discussed.
Finally, conclusions are provided in section-5.

\section{Field Equations \& Wormhole Geometry}
The static and spherically symmetric  metric defining the wormhole structure  is
\begin{equation}\label{metric}
ds^2=-e^{2\Phi(r)}dt^2+\frac{dr^2}{1-b(r)/r} + r^2(d\theta^2+\sin^2\theta^2\phi^2).
\end{equation}
The  function $\Phi (r)$ determines the gravitational redshift, hence it is called redshift function.
The wormhole solutions must satisfy Einstein's field equations and must possess a throat that joins two regions of universe which are asymptotically flat. For a traversable wormhole, event horizon should not be present and the effect of tidal gravitational forces should be very small on a traveler.

The functions $\Phi(r)$ and $b(r)$ are the functions of  radial coordinate $r$, which is a non-decreasing function. Its minimum value is $r_0$, radius of the throat, and maximum value is $+\infty$.  The function $b(r)$ determines the shape of wormhole, hence it is called as shape function. The existence of wormhole solutions demands the satisfaction of following conditions:
(i) $b(r_0)=r_0$, (ii) $\frac{b(r)-b'(r)r}{b^2}>0$, (iii) $b'(r_0)-1\leq 0$, (iv) $\frac{b(r)}{r}<1$ for $r>r_0$ and (v) $\frac{b(r)}{r}\rightarrow 0$ as $r\rightarrow\infty$. For simplicity, the redshift function is assumed as a constant.

Morris \& Thorne \cite{morris1} introduced traversable wormholes in the framework of Einstein's general theory of relativity. The $f(R)$  theory of gravity is a generalization of  Einstein's theory of relativity which  replaces the  gravitational action $R$ with a general function $f(R)$ of $R$. Thus, the  gravitational action for $f(R)$ theory of gravity is defined as
\begin{equation}\label{action}
S_G=\dfrac{1}{2k}\int[f(R) + L_m]\sqrt{-g}d^4x,
\end{equation}
where $k=8\pi G$, $L_m$ and $g$ stand for the  matter Lagrangian density and  the  determinant of the metric $g_{\mu\nu}$ respectively. For simplicity $k$ is taken as unity.\\

Variation of Eq.(\ref{action}) with respect to the metric $g_{\mu\nu}$ gives the field equations as
\begin{equation}\label{fe}
FR_{\mu\nu} -\dfrac{1}{2}fg_{\mu\nu}-\triangledown_\mu\triangledown_\nu F+\square Fg_{\mu\nu}= T_{\mu\nu}^m,
\end{equation}	
where $R_{\mu\nu}$ and $R$ denote Ricci tensor and curvature scalar respectively and $F=\frac{df}{dR}$. The contraction of \ref{fe}, gives
\begin{equation}\label{trace}
FR-2f+3\square F=T,
\end{equation}
where $T=T^{\mu}_{\mu}$ is the trace of the stress energy tensor.

From Eqs. \ref{fe} \& \ref{trace}, the effective field equations are obtained as
\begin{equation}
G_{\mu\nu}\equiv R_{\mu\nu}-\frac{1}{2}Rg_{\mu\nu}=T_{\mu\nu}^{eff},
\end{equation}
where $T_{\mu\nu}^{eff}=T_{\mu\nu}^{c}+T_{\mu\nu}^{m}/F$ and $T_{\mu\nu}^{c}=\frac{1}{F}[\triangledown_\mu\triangledown_\nu F-\frac{1}{4}g_{\mu\nu}(FR+\square F+T)]$.
The energy momentum tensor for the matter source of the wormholes is $T_{\mu\nu}=\frac{\partial L_m}{\partial g^{\mu\nu}}$, which is defined as
\begin{equation}
T_{\mu\nu} = (\rho + p_t)u_\mu u_\nu - p_tg_{\mu\nu}+(p_r-p_t)X_\mu X_\nu,
\end{equation}	
such that
\begin{equation}
u^{\mu}u_\mu=-1 \mbox{ and } X^{\mu}X_\mu=1,
\end{equation}

where $\rho$,  $p_t$ and $p_r$  stand for the energy density, tangential pressure and radial pressure respectively.

The  Ricci scalar $R$ given by $R=\frac{2b'(r)}{r^2}$ and Einstein's field equations for the metric \ref{metric} in  $f(R)$ gravity are obtained as:
\begin{equation}\label{6}
\rho=\frac{Fb'(r)}{r^2}-H
\end{equation}
\begin{equation}\label{7}
p_r=-\frac{b(r)F}{r^3}-\Bigg(1-\frac{b(r)}{r}\Bigg)\Bigg[F''+\frac{F'(rb'(r)-b(r))}{2r^2\Big(1-\frac{b(r)}{r}\Big)}\Bigg]+H
\end{equation}
\begin{equation}\label{8}
p_t=\frac{F(b(r)-rb'(r))}{2r^3}-\frac{F'}{r}\Bigg(1-\frac{b(r)}{r}\Bigg)+H,
\end{equation}

where $H=\frac{1}{4}(FR+\square F+T)$ and prime upon a function denotes the derivative of that function with respect to  radial coordinate $r$.
%

\section{Wormhole Solutions}

Amendola et al. \cite{Amendola} investigated the conditions for $f(R)$ dark energy models to be cosmologically viable and used geometrical approach to explore the cosmological nature of $f(R)$ models.  They derived autonomous equations for arbitrary $f(R)$ models, determined all fixed points for such system and examined the stability of these points to study the cosmological evolution. They drew $m(r)$ curves in the $rm-$plane, where $m=\frac{R\frac{dF}{dR}}{F}$, $r=-\frac{RF}{f}$ \& $F=\frac{df}{dR}$, and categorize $f(R)$ models into four classes and studied their cosmological viability. They obtained the $f(R)$ model with
\begin{equation}\label{model}
f(R)=R-\mu R_c\Big(\frac{R}{R_c}\Big)^p,
\end{equation}
 where $\mu, R_c>0$  and $0<p<1$, belonging to Class II and satisfying all the requirements to lead towards an acceptable cosmology.

A viable $f(R)$ dark energy model should satisfy the following conditions: (i) $F>0$ for $R\geq R_0$, where $R_0$ is the Ricci scalar at the present epoch, if the final attractor is a de Sitter point with the Ricci scalar $R_1$, then $F> 0$ for $R \geq R_1$; (ii) $\frac{dF}{dR}>0$ for $R\geq R_0$; (iii) $F\rightarrow R-\Lambda$ for $R>>R_0$ and (iv) $0<m<1$ at $r=-2$.
Model (\ref{model}) satisfies these four conditions. To satisfy condition (iii), the power $p$ in model (\ref{model}) should be close to zero.
The experimental bound on model  (\ref{model}) is obtained as $\frac{p}{2-p}\Big(\frac{R_1}{\rho_B}\Big)^{1-p}<1.5\times 10^{-15}$, where  $\rho_B$ is the density outside the body. Assuming $R_1=10^{-29}$ g/cm$^3$ \& $\rho_B=10^{-24}$ g/cm$^3$, the constraint on $p$ is obtained as $p<3\times 10^{-10}$ which shows a very small deviation from $\Lambda$CDM model \cite{Dark}. These facts have motivated to choose power law model \ref{model}. 
 In this paper, this $f(R)$ model is considered to explore the geometry of wormholes. The field equations are solved and the energy condition terms are derived which are as follows:

\begin{eqnarray}
\rho&=&\frac{1}{(r+(r+1) \log (r+1))^4}\Bigg[R_c^2 \mu  2^{p-2} (p-1) p \log
({r_0}+1) \Big((r+1) \log ^2(r+1) \Big(r \Big(4 p (r+1)\nonumber\\
&+&r^2-4 r-9\Big)+2 (p-2) (r+1)^3 \log ({r_0}+1)\Big)+r^2 \Big(2 \Big(p-r^2-3 r-2\Big) \log ({r_0}+1)+r\Big)\nonumber\\
&+&r \log (r+1) \Big(r \Big(-2 p+r^2+5 r+4\Big)-2 (r+1) \Big(2 p (r+1)+r^2-r-4\Big) \log ({r_0}+1)\Big)\nonumber\\
&-&2 (p-2) (r+1)^4 \log ^3(r+1)\Big) \Bigg(\frac{r+(r+1) \log (r+1)}{R_c r (r+1) \log ({r_0}+1)}\Big)^{p+1}\Bigg]+\frac{1}{r^2 (r+1) \log ({r_0}+1)}\Bigg[(r\nonumber\\
&+&(r+1) \log (r+1)) \Big(1-\mu  2^{p-1} p \Big(\frac{r+(r+1) \log (r+1)}{R_c r (r+1) \log ({r_0}+1)}\Big)^{p-1}\Big)\Bigg]+R_c \mu  (-2^{p-1})\nonumber\\ &\times&\Big(\frac{r+(r+1) \log (r+1)}{R_c r (r+1) \log ({r_0}+1)}\Big)^p-\frac{(r+(r+1) \log (r+1)) \Big(1-\mu  2^{p-1} p \Big(\frac{r+(r+1) \log (r+1)}{R_c r (r+1) \log ({r_0}+1)}\Big)^{p-1}\Big)}{r (r+1) \log ({r_0}+1)}\nonumber\\
&+&\frac{\frac{1}{r+1}+\frac{\log (r+1)}{r}}{\log ({r_0}+1)}
\end{eqnarray}

\begin{eqnarray}
p_r&=&-\frac{1}{2 r^2 (r+(r+1) \log (r+1))^2 \log ({r_0}+1)}\Bigg[-r \log (r+1) \left(-R_c \mu  2^p r \left(-2 p^2+p \left(2 r^2+r\right.\right.\right.\nonumber\\
&+&1\left.\left.\left.\right)-2 r (r+1)\right) \log ({r_0}+1) \left(\frac{r+(r+1) \log (r+1)}{R_c r (r+1) \log ({r_0}+1)}\right)^p+R_c \mu  2^{p+1} (p-1) p (r+1)^2 \right.\nonumber\\
&\times&\left.\log ^2({r_0}+1) \left(\frac{r+(r+1) \log (r+1)}{R_c r (r+1) \log ({r_0}+1)}\right)^p-2 r\right)+r (r+1) \log ^2(r+1) \left(R_c \mu  2^p (r+1)\right.\nonumber\\
&\times& \left.\left(2 p^2+p (r-3)-r\right) \log ({r_0}+1) \left(\frac{r+(r+1) \log (r+1)}{R_c r (r+1) \log ({r_0}+1)}\right)^p+4\right)+R_c \mu  2^p (p-1) r^2\nonumber\\
&\times& \log ({r_0}+1) \left(2 p \log ({r_0}+1)+r^2\right) \left(\frac{r+(r+1) \log (r+1)}{R_c r (r+1) \log ({r_0}+1)}\right)^p+2 (r+1)^2 \log ^3(r+1)\Bigg]
\end{eqnarray}

\begin{eqnarray}
p_t&=&\frac{1}{4 r (r+1) \log ({r_0}+1)}\Bigg[-\frac{1}{4 r (r+1) \log ({r_0}+1)}\Bigg(R_c \mu  2^p (p-1) p \log ({r_0}+1) \left((r+1)\right.\nonumber\\
&\times& \left.\log ^2(r+1) \left(r \left(4 p (r+1)+r^2-4 r-9\right)+2 (p-2) (r+1)^3 \log ({r_0}+1)\right)+r^2 \left(2 \left(p-r^2\right.\right.\right.\nonumber\\
&-&\left.\left.\left.3 r-2\right) \log ({r_0}+1)+r\right)+r \log (r+1) \left(r \left(-2 p+r^2+5 r+4\right)-2 (r+1) \left(2 p (r+1)\right.\right.\right.\nonumber\\
&+&\left.\left.\left.r^2-r-4\right) \log ({r_0}+1)\right)-2 (p-2) (r+1)^4 \log ^3(r+1)\right) \left(\frac{r+(r+1) \log (r+1)}{R_c r (r+1) \log ({r_0}+1)}\right)^p\Bigg)\nonumber
\end{eqnarray}

\begin{eqnarray}
&+&\mu  2^p p \left(\frac{r+(r+1) \log (r+1)}{R_c r (r+1) \log ({r_0}+1)}\right)^{p-1}+\frac{1}{r (r+(r+1) \log (r+1))}\Bigg(\mu  2^{p+1} (p-1) p \left((r+1)^2\right. \nonumber\\
&\times&\left.\log (r+1)-r\right) (\log (r+1)-\log ({r_0}+1)) \left(\frac{r+(r+1) \log (r+1)}{R_c r (r+1) \log ({r_0}+1)}\right)^{p-1}\Bigg)+2 r \left(R_c \mu  2^p\right. \nonumber\\
&\times&\left.(r+1) \log ({r_0}+1) \left(\frac{r+(r+1) \log (r+1)}{R_c r (r+1) \log ({r_0}+1)}\right)^p-2\right)+4 (r+(r+1) \log (r+1)) \nonumber\\
&\times&\left(1-\mu  2^{p-1} p \left(\frac{r+(r+1) \log (r+1)}{R_c r (r+1) \log ({r_0}+1)}\right)^{p-1}\right)-4 (r+1) \log (r+1)-2\Bigg]
\end{eqnarray}

\begin{eqnarray}
\rho+p_r&=&\frac{1}{(r+(r+1) \log (r+1))^4}\Bigg[R_c^2 \mu  2^{p-2} (p-1) p \log
({r_0}+1) \Big((r+1) \log ^2(r+1) \Big(r \Big(4 p (r+1)\nonumber\\
&+&r^2-4 r-9\Big)+2 (p-2) (r+1)^3 \log ({r_0}+1)\Big)+r^2 \Big(2 \Big(p-r^2-3 r-2\Big) \log ({r_0}+1)+r\Big)\nonumber\\
&+&r \log (r+1) \Big(r \Big(-2 p+r^2+5 r+4\Big)-2 (r+1) \Big(2 p (r+1)+r^2-r-4\Big) \log ({r_0}+1)\Big)\nonumber\\
&-&2 (p-2) (r+1)^4 \log ^3(r+1)\Big) \Bigg(\frac{r+(r+1) \log (r+1)}{R_c r (r+1) \log ({r_0}+1)}\Big)^{p+1}\Bigg]+\frac{1}{r^2 (r+1) \log ({r_0}+1)}\Bigg[(r\nonumber\\
&+&(r+1) \log (r+1)) \Big(1-\mu  2^{p-1} p \Big(\frac{r+(r+1) \log (r+1)}{R_c r (r+1) \log ({r_0}+1)}\Big)^{p-1}\Big)\Bigg]+R_c \mu  (-2^{p-1})\nonumber\\ &\times&\Big(\frac{r+(r+1) \log (r+1)}{R_c r (r+1) \log ({r_0}+1)}\Big)^p-\frac{1}{r (r+1) \log ({r_0}+1)}\Bigg[(r+(r+1) \log (r+1)) \Big(1-\mu  2^{p-1} p\nonumber\\
&\times& \Big(\frac{r+(r+1) \log (r+1)}{R_c r (r+1) \log ({r_0}+1)}\Big)^{p-1}\Big)\Bigg]+\frac{\frac{1}{r+1}+\frac{\log (r+1)}{r}}{\log ({r_0}+1)}\nonumber
\end{eqnarray}
\begin{eqnarray}
&-&\frac{1}{2 r^2 (r+(r+1) \log (r+1))^2 \log ({r_0}+1)}\Bigg[-r \log (r+1) \left(-R_c \mu  2^p r \left(-2 p^2+p \left(2 r^2+r\right.\right.\right.\nonumber\\
&+&1\left.\left.\left.\right)-2 r (r+1)\right) \log ({r_0}+1) \left(\frac{r+(r+1) \log (r+1)}{R_c r (r+1) \log ({r_0}+1)}\right)^p+R_c \mu  2^{p+1} (p-1) p (r+1)^2 \right.\nonumber\\
&\times&\left.\log ^2({r_0}+1) \left(\frac{r+(r+1) \log (r+1)}{R_c r (r+1) \log ({r_0}+1)}\right)^p-2 r\right)+r (r+1) \log ^2(r+1) \left(R_c \mu  2^p (r+1)\right.\nonumber\\
&\times& \left.\left(2 p^2+p (r-3)-r\right) \log ({r_0}+1) \left(\frac{r+(r+1) \log (r+1)}{R_c r (r+1) \log ({r_0}+1)}\right)^p+4\right)+R_c \mu  2^p (p-1) r^2\nonumber\\
&\times& \log ({r_0}+1) \left(2 p \log ({r_0}+1)+r^2\right) \left(\frac{r+(r+1) \log (r+1)}{R_c r (r+1) \log ({r_0}+1)}\right)^p+2 (r+1)^2 \log ^3(r+1)\Bigg]
\end{eqnarray}

\begin{eqnarray}
\rho+p_t&=&\frac{1}{(r+(r+1) \log (r+1))^4}\Bigg[R_c^2 \mu  2^{p-2} (p-1) p \log
({r_0}+1) \Big((r+1) \log ^2(r+1) \Big(r \Big(4 p (r+1)\nonumber\\
&+&r^2-4 r-9\Big)+2 (p-2) (r+1)^3 \log ({r_0}+1)\Big)+r^2 \Big(2 \Big(p-r^2-3 r-2\Big) \log ({r_0}+1)+r\Big)\nonumber\\
&+&r \log (r+1) \Big(r \Big(-2 p+r^2+5 r+4\Big)-2 (r+1) \Big(2 p (r+1)+r^2-r-4\Big) \log ({r_0}+1)\Big)\nonumber\\
&-&2 (p-2) (r+1)^4 \log ^3(r+1)\Big) \Bigg(\frac{r+(r+1) \log (r+1)}{R_c r (r+1) \log ({r_0}+1)}\Big)^{p+1}\Bigg]+\frac{1}{r^2 (r+1) \log ({r_0}+1)}\Bigg[(r\nonumber\\
&+&(r+1) \log (r+1)) \Big(1-\mu  2^{p-1} p \Big(\frac{r+(r+1) \log (r+1)}{R_c r (r+1) \log ({r_0}+1)}\Big)^{p-1}\Big)\Bigg]+R_c \mu  (-2^{p-1})\nonumber\\ &\times&\Big(\frac{r+(r+1) \log (r+1)}{R_c r (r+1) \log ({r_0}+1)}\Big)^p-\frac{1}{r (r+1) \log ({r_0}+1)}\Bigg[(r+(r+1) \log (r+1)) \Big(1-\mu  2^{p-1} p\nonumber\\
&\times& \Big(\frac{r+(r+1) \log (r+1)}{R_c r (r+1) \log ({r_0}+1)}\Big)^{p-1}\Big)\Bigg]+\frac{\frac{1}{r+1}+\frac{\log (r+1)}{r}}{\log ({r_0}+1)}\nonumber\\
&+&\frac{1}{4 r (r+1) \log ({r_0}+1)}\Bigg[-\frac{1}{4 r (r+1) \log ({r_0}+1)}\Bigg(R_c \mu  2^p (p-1) p \log ({r_0}+1) \left((r+1)\right.\nonumber\\
&\times& \left.\log ^2(r+1) \left(r \left(4 p (r+1)+r^2-4 r-9\right)+2 (p-2) (r+1)^3 \log ({r_0}+1)\right)+r^2 \left(2 \left(p-r^2\right.\right.\right.\nonumber\\
&-&\left.\left.\left.3 r-2\right) \log ({r_0}+1)+r\right)+r \log (r+1) \left(r \left(-2 p+r^2+5 r+4\right)-2 (r+1) \left(2 p (r+1)\right.\right.\right.\nonumber\\
&+&\left.\left.\left.r^2-r-4\right) \log ({r_0}+1)\right)-2 (p-2) (r+1)^4 \log ^3(r+1)\right) \left(\frac{r+(r+1) \log (r+1)}{R_c r (r+1) \log ({r_0}+1)}\right)^p\Bigg)\nonumber\\
&+&\mu  2^p p \left(\frac{r+(r+1) \log (r+1)}{R_c r (r+1) \log ({r_0}+1)}\right)^{p-1}+\frac{1}{r (r+(r+1) \log (r+1))}\Bigg(\mu  2^{p+1} (p-1) p \left((r+1)^2\right. \nonumber\\
&\times&\left.\log (r+1)-r\right) (\log (r+1)-\log ({r_0}+1)) \left(\frac{r+(r+1) \log (r+1)}{R_c r (r+1) \log ({r_0}+1)}\right)^{p-1}\Bigg)+2 r \left(R_c \mu  2^p\right. \nonumber\\
&\times&\left.(r+1) \log ({r_0}+1) \left(\frac{r+(r+1) \log (r+1)}{R_c r (r+1) \log ({r_0}+1)}\right)^p-2\right)+4 (r+(r+1) \log (r+1)) \nonumber\\
&\times&\left(1-\mu  2^{p-1} p \left(\frac{r+(r+1) \log (r+1)}{R_c r (r+1) \log ({r_0}+1)}\right)^{p-1}\right)-4 (r+1) \log (r+1)-2\Bigg]
\end{eqnarray}

\begin{eqnarray}
\rho-|p_r|&=&\frac{1}{(r+(r+1) \log (r+1))^4}\Bigg[R_c^2 \mu  2^{p-2} (p-1) p \log
({r_0}+1) \Big((r+1) \log ^2(r+1) \Big(r \Big(4 p (r+1)\nonumber\\
&+&r^2-4 r-9\Big)+2 (p-2) (r+1)^3 \log ({r_0}+1)\Big)+r^2 \Big(2 \Big(p-r^2-3 r-2\Big) \log ({r_0}+1)+r\Big)\nonumber
\end{eqnarray}

\begin{eqnarray}
&+&r \log (r+1) \Big(r \Big(-2 p+r^2+5 r+4\Big)-2 (r+1) \Big(2 p (r+1)+r^2-r-4\Big) \log ({r_0}+1)\Big)\nonumber\\
&-&2 (p-2) (r+1)^4 \log ^3(r+1)\Big) \Bigg(\frac{r+(r+1) \log (r+1)}{R_c r (r+1) \log ({r_0}+1)}\Big)^{p+1}\Bigg]+\frac{1}{r^2 (r+1) \log ({r_0}+1)}\Bigg[(r\nonumber\\
&+&(r+1) \log (r+1)) \Big(1-\mu  2^{p-1} p \Big(\frac{r+(r+1) \log (r+1)}{R_c r (r+1) \log ({r_0}+1)}\Big)^{p-1}\Big)\Bigg]+R_c \mu  (-2^{p-1})\nonumber\\ &\times&\Big(\frac{r+(r+1) \log (r+1)}{R_c r (r+1) \log ({r_0}+1)}\Big)^p-\frac{1}{r (r+1) \log ({r_0}+1)}\Bigg[(r+(r+1) \log (r+1)) \Big(1-\mu  2^{p-1} p\nonumber\\
&\times& \Big(\frac{r+(r+1) \log (r+1)}{R_c r (r+1) \log ({r_0}+1)}\Big)^{p-1}\Big)\Bigg]+\frac{\frac{1}{r+1}+\frac{\log (r+1)}{r}}{\log ({r_0}+1)}\nonumber\\
&-&\Bigg|-\frac{1}{2 r^2 (r+(r+1) \log (r+1))^2 \log ({r_0}+1)}\Bigg[-r \log (r+1) \left(-R_c \mu  2^p r \left(-2 p^2+p \left(2 r^2+r\right.\right.\right.\nonumber\\
&+&1\left.\left.\left.\right)-2 r (r+1)\right) \log ({r_0}+1) \left(\frac{r+(r+1) \log (r+1)}{R_c r (r+1) \log ({r_0}+1)}\right)^p+R_c \mu  2^{p+1} (p-1) p (r+1)^2 \right.\nonumber\\
&\times&\left.\log ^2({r_0}+1) \left(\frac{r+(r+1) \log (r+1)}{R_c r (r+1) \log ({r_0}+1)}\right)^p-2 r\right)+r (r+1) \log ^2(r+1) \left(R_c \mu  2^p (r+1)\right.\nonumber\\
&\times& \left.\left(2 p^2+p (r-3)-r\right) \log ({r_0}+1) \left(\frac{r+(r+1) \log (r+1)}{R_c r (r+1) \log ({r_0}+1)}\right)^p+4\right)+R_c \mu  2^p (p-1) r^2\nonumber\\
&\times& \log ({r_0}+1) \left(2 p \log ({r_0}+1)+r^2\right) \left(\frac{r+(r+1) \log (r+1)}{R_c r (r+1) \log ({r_0}+1)}\right)^p+2 (r+1)^2 \log ^3(r+1)\Bigg]
\Bigg|
\end{eqnarray}

\begin{eqnarray}
\rho-|p_t|&=&\frac{1}{(r+(r+1) \log (r+1))^4}\Bigg[R_c^2 \mu  2^{p-2} (p-1) p \log
({r_0}+1) \Big((r+1) \log ^2(r+1) \Big(r \Big(4 p (r+1)\nonumber\\
&+&r^2-4 r-9\Big)+2 (p-2) (r+1)^3 \log ({r_0}+1)\Big)+r^2 \Big(2 \Big(p-r^2-3 r-2\Big) \log ({r_0}+1)+r\Big)\nonumber\\
&+&r \log (r+1) \Big(r \Big(-2 p+r^2+5 r+4\Big)-2 (r+1) \Big(2 p (r+1)+r^2-r-4\Big) \log ({r_0}+1)\Big)\nonumber\\
&-&2 (p-2) (r+1)^4 \log ^3(r+1)\Big) \Bigg(\frac{r+(r+1) \log (r+1)}{R_c r (r+1) \log ({r_0}+1)}\Big)^{p+1}\Bigg]+\frac{1}{r^2 (r+1) \log ({r_0}+1)}\Bigg[(r\nonumber\\
&+&(r+1) \log (r+1)) \Big(1-\mu  2^{p-1} p \Big(\frac{r+(r+1) \log (r+1)}{R_c r (r+1) \log ({r_0}+1)}\Big)^{p-1}\Big)\Bigg]+R_c \mu  (-2^{p-1})\nonumber\\ &\times&\Big(\frac{r+(r+1) \log (r+1)}{R_c r (r+1) \log ({r_0}+1)}\Big)^p-\frac{1}{r (r+1) \log ({r_0}+1)}\Bigg[(r+(r+1) \log (r+1)) \Big(1-\mu  2^{p-1} p\nonumber
\end{eqnarray}

\begin{eqnarray}
&\times& \Big(\frac{r+(r+1) \log (r+1)}{R_c r (r+1) \log ({r_0}+1)}\Big)^{p-1}\Big)\Bigg]+\frac{\frac{1}{r+1}+\frac{\log (r+1)}{r}}{\log ({r_0}+1)}\nonumber\\
&-&\Bigg|\frac{1}{4 r (r+1) \log ({r_0}+1)}\Bigg[-\frac{1}{4 r (r+1) \log ({r_0}+1)}\Bigg(R_c \mu  2^p (p-1) p \log ({r_0}+1) \left((r+1)\right.\nonumber\\
&\times& \left.\log ^2(r+1) \left(r \left(4 p (r+1)+r^2-4 r-9\right)+2 (p-2) (r+1)^3 \log ({r_0}+1)\right)+r^2 \left(2 \left(p-r^2\right.\right.\right.\nonumber\\
&-&\left.\left.\left.3 r-2\right) \log ({r_0}+1)+r\right)+r \log (r+1) \left(r \left(-2 p+r^2+5 r+4\right)-2 (r+1) \left(2 p (r+1)\right.\right.\right.\nonumber\\
&+&\left.\left.\left.r^2-r-4\right) \log ({r_0}+1)\right)-2 (p-2) (r+1)^4 \log ^3(r+1)\right) \left(\frac{r+(r+1) \log (r+1)}{R_c r (r+1) \log ({r_0}+1)}\right)^p\Bigg)\nonumber\\
&+&\mu  2^p p \left(\frac{r+(r+1) \log (r+1)}{R_c r (r+1) \log ({r_0}+1)}\right)^{p-1}+\frac{1}{r (r+(r+1) \log (r+1))}\Bigg(\mu  2^{p+1} (p-1) p \left((r+1)^2\right. \nonumber\\
&\times&\left.\log (r+1)-r\right) (\log (r+1)-\log ({r_0}+1)) \left(\frac{r+(r+1) \log (r+1)}{R_c r (r+1) \log ({r_0}+1)}\right)^{p-1}\Bigg)+2 r \left(R_c \mu  2^p\right. \nonumber\\
&\times&\left.(r+1) \log ({r_0}+1) \left(\frac{r+(r+1) \log (r+1)}{R_c r (r+1) \log ({r_0}+1)}\right)^p-2\right)+4 (r+(r+1) \log (r+1)) \nonumber\\
&\times&\left(1-\mu  2^{p-1} p \left(\frac{r+(r+1) \log (r+1)}{R_c r (r+1) \log ({r_0}+1)}\right)^{p-1}\right)-4 (r+1) \log (r+1)-2\Bigg]\Bigg|
\end{eqnarray}

\begin{figure}
	\centering
	\subfigure[$\rho$]{\includegraphics[scale=.46]{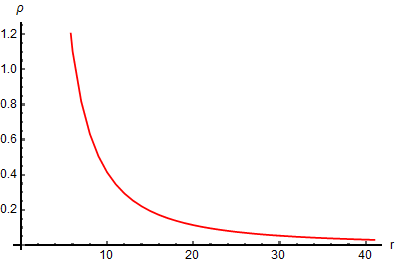}}\hspace{.1cm}
	\subfigure[$\rho+p_r$]{\includegraphics[scale=.46]{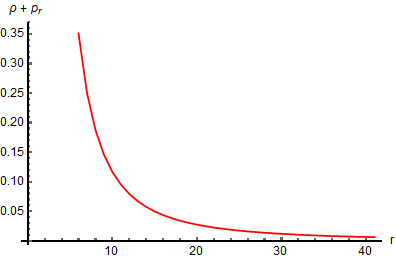}}\hspace{.1cm}
	\subfigure[$\rho+p_t$]{\includegraphics[scale=.46]{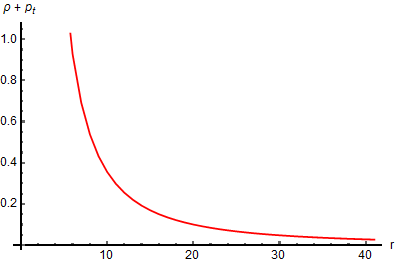}}\hspace{.1cm}
	\subfigure[$\rho-|p_r|$]{\includegraphics[scale=.46]{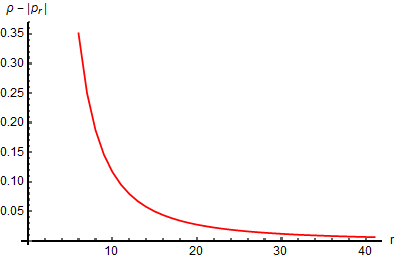}}\hspace{.1cm}
	\subfigure[$\rho-|p_t|$]{\includegraphics[scale=.46]{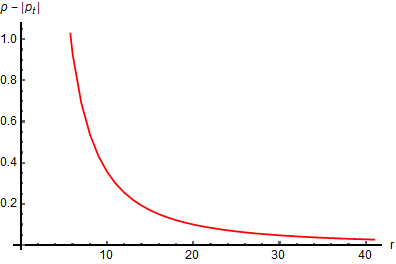}}\hspace{.1cm}
	\subfigure[$\rho+p_r+2p_t$]{\includegraphics[scale=.46]{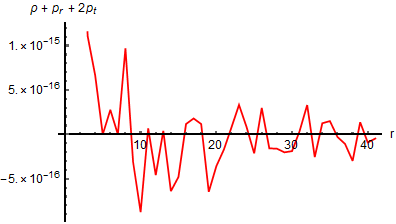}}\hspace{.1cm}
	\subfigure[$\triangle$]{\includegraphics[scale=.46]{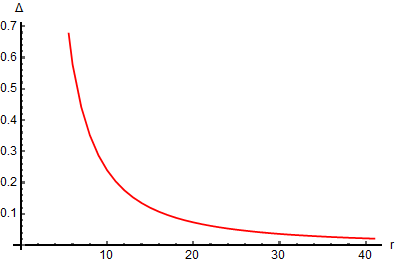}}\hspace{.1cm}
	\subfigure[$\omega$]{\includegraphics[scale=.46]{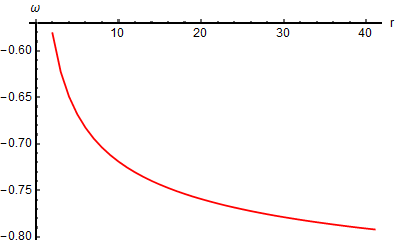}}\hspace{.1cm}
	\caption{Case 2: Plots for Density, NEC, DEC, $\triangle$ \& $\omega$ with $f(R) = R + \alpha R^m$}
\end{figure}
\section{Results \& Discussion}
The wormhole metric is dependent on shape function defined in terms of radial coordinate. It plays a significant role in determining the characteristics of wormhole structure. In literature, many shape functions are defined and used to study wormholes. Godani and Samanta \cite{godani} defined a logarithmic shape function $b(r)=\frac{r\log(r+1)}{\log(r_0+1)}$ and explored traversable wormholes in $f(R)$ gravity. In this work, this shape function is considered to investigate the wormhole structure in modified $f(R)$ theory of gravity with the function $f(R)=R-\mu R_c\Big(\frac{R}{R_c}\Big)^p$, where $\mu$, $R_c>0$ and $0<p<1$. The obtained results are discussed below:

It is observed that the energy density is positive for every value of radial coordinate $r$ and constant $\mu$, if $R_c<<1$. Therefore, we have taken $R_c=10^{-29}$ to obtain the results. Subsequently, the energy density is obtained to be a positive and decreasing function of $r$ (Fig. 1(a)).  For any null vector, the null energy condition (NEC) is defined as
$NEC\Leftrightarrow T_{\mu\nu}k^{\mu}k^{\nu}\ge 0$, in terms of the principal pressures $NEC\Leftrightarrow ~~ \forall i, ~\rho+p_{i}\ge 0$. So, we have examined $\rho+p_r$ and $\rho+p_t$. It is found that both NEC terms $\rho+p_r$ and $\rho+p_t$ are positive functions (Figs. 1(b), 1(c)) and hence the null energy condition is satisfied. Now, for a timelike vector, the weak energy condition (WEC) is defined as $WEC\Leftrightarrow T_{\mu\nu}V^{\mu}V^{\nu}\ge 0$. In terms of the principal pressures, it is defined as $WEC\Leftrightarrow \rho\ge 0;$ and $\forall i,  ~~ \rho+p_{i}\ge 0$. Since the energy density is positive and NEC is satisfied, therefore WEC is also satisfied. Further,  for any timelike vector, the dominant energy condition (DEC) is defined as $DEC\Leftrightarrow T_{\mu\nu}V^{\mu}V^{\nu}\ge 0$, and
$T_{\mu\nu}V^{\mu}$ is not spacelike. In terms of the principal pressures,
$DEC\Leftrightarrow \rho\ge 0;$ and $\forall i, ~ p_i\in [-\rho, ~+\rho]$. For the model considered we have checked the terms $\rho-|p_r|$ \& $\rho-|p_t|$. Both terms are observed to be positive functions of $r$ throughout (Figs. 1(d), 1(e)). This shows the satisfaction of DEC. For a timelike vector, the strong energy condition (SEC) is defined as $SEC\Leftrightarrow (T_{\mu\nu}-\frac{T}{2}g_{\mu\nu})V^{\mu}V^{\nu}\ge 0$, where $T$ is the trace of the stress-energy tensor. The strong energy condition implies the null energy condition but it does not imply, in general, the weak energy condition. In terms of the principal pressures, $T=-\rho+\sum_j{p_j}$ and $SEC\Leftrightarrow \forall j, ~ \rho+p_j\geq 0, ~ \rho+\sum_j{p_j}\geq 0$. For the wormhole model undertaken, we have analyzed that SEC term $\rho+p_r+2p_t$ oscillates between positive and negative values (Fig. 1(f)). So, SEC is violated. Thus, all energy conditions NEC, WEC \& DEC except SEC are satisfied throughout.
The anisotropy parameter is defined as $\triangle=p_t-p_r$. The positive value of $\triangle$ represents repulsive geometry, while its negative value represents attractive nature of geometry. For our model, the value of anisotropy parameter is obtained to be positive which shows the repulsive nature of geometry inside the wormhole (Fig. 1(g)). Further, the equation of state parameter in terms of radial pressure is defined as $\omega=\frac{p_r}{\rho}$. The value of $\omega$ tells about the type of fluid filled.   For our model, its value is obtained to lie between -1 and 0 which shows the presence of non-phantom fluid filled inside the structure of wormhole (Fig. 1(h)). Thus, it is analyzed that the wormhole solutions exists without presence of exotic matter. This has become possible because of a suitable choice of the shape function and $f(R)$ function.

\section{Conclusion}
Amendola et al. \cite{Amendola} obtained the conditions for $f(R)$ dark energy models to be cosmologically viable and examined  the cosmological behavior of various $f(R)$ models. The $f(R)$ model with $f(R)=R-\mu R_c\Big(\frac{R}{R_c}\Big)^p$, where $\mu, R_c>0$  and $0<p<1$, is one of them. They found this model to fulfill all the requirements to be a cosmologically acceptable model. In the present work, this $f(R)$ model is taken into account to study wormhole metric in $f(R)$ gravity.  The energy conditions are examined and geometric nature is analyzed. The null, weak and dominated energy conditions are found to agree everywhere. This depicts the existence of wormhole solutions without requirement of non-exotic matter. The geometric structure of wormhole is inspected to be repulsive in nature and filled with non-phantom fluid.   Thus, the results obtained strongly support the presence of non-exotic matter and consequently, ensure the existence of wormhole geometry. Hence, from this work we may conclude that traversable wormholes are possible without the support of exotic matter.    \\

\textbf{Acknowledgement:} The authors are very much thankful to the anonymous reviewer for the constructive comments for improvement of the work.

\end{document}